\documentclass[letterpaper, 10 pt, conference]{ieeeconf}  

\IEEEoverridecommandlockouts                              
\usepackage{array}
\newcolumntype{P}[1]{>{\centering\arraybackslash}p{#1}}
\usepackage{color}
\usepackage{multirow}
\usepackage{amsmath}
\usepackage{algorithm}
\usepackage[noend]{algpseudocode}
\usepackage{mathtools}
\usepackage{svg}
\usepackage{amsmath}

\usepackage{graphicx} 
\usepackage{array}
\newcolumntype{P}[1]{>{\centering\arraybackslash}p{#1}}
\usepackage{color}
\usepackage{multirow}
\usepackage{amsmath}
\usepackage{algorithm}
\usepackage[noend]{algpseudocode}
\usepackage{mathtools}
\usepackage{svg}
\usepackage{amsmath}

\usepackage{color}
\usepackage{multirow}
\usepackage{amsmath}
\usepackage{algorithm}
\usepackage[noend]{algpseudocode}
\usepackage{mathtools}
\usepackage{svg}
\usepackage{amsmath}

\title{\LARGE \bf
Design Space Exploration of Algorithmic Multi-port Memory for High-Performance Application-Specific Accelerators}

\author{Khushal Sethi \\
Department of Electrical Engineering\\
Indian Institute of Technology, Delhi \\
{\tt\small ee1160556@iitd.ac.in}}

\begin{document}

\maketitle
\thispagestyle{empty}
\pagestyle{empty}

\begin{abstract}
Memory load/store instructions consume an important part in execution time and energy consumption in domain-specific accelerators. For designing highly parallel systems, available parallelism at each granularity is extracted from the workloads.
The maximal use of parallelism at each granularity in these high-performance designs requires the utilization of multi-port memories.
Currently, true multiport designs are less popular because there is no inherent EDA support for multiport memory beyond 2-ports, utilizing more ports requires circuit-level implementation and hence a very high design time.
In this work, we present a framework for Design Space Exploration of Algorithmic Multi-Port Memories (AMM) in ASICs. We study different AMM designs in the literature, discuss how we incorporate them in the Pre-RTL Aladdin Framework with different memory depth, port configurations and banking structures. From our analysis on selected applications from the MachSuite (accelerator benchmark suite), we understand and quantify the potential use of AMMs (as true multiport memories) for high performance in applications with low spatial locality in memory access patterns. 
\end{abstract}
\section{INTRODUCTION}
There has been a rise in the popularity of hardware accelerators, especially in the mobile domain (SoCs). This includes designing domain-specific accelerators/implementations for machine learning \cite{radway2021future, bashir2019power, sethi2021efficient, ji2020reconfigurable,ji2020compacc, sethi2020nv, sethi2020design, sethi2018low, sethi2019optimized, sethi2022dragon}, graph processing \cite{gui2019survey}, DNA-Alignment algorithms \cite{kim2018grim}, Neuromorphic algorithms \cite{sethi2019optimized} etc.  

Memory design is a crucial component of hardware accelerators, where designers require stringent architectures for memory accesses to gain high-performance. A high-performance spatial accelerator (as shown in Fig 1) can efficiently exploit the parallelism of the data-flow graph, and execute multiple operations in parallel.  Hence, multiport memories may be required to support single-cycle access in hardware. 

However, there is currently no inherent EDA support for multiport memories beyond 2-ports, hence for utilizing more ports, a circuit-level implementation has to be done from scratch. Algorithmic Multiported Memory (AMM) allows us to utilize multiple ports starting from the same circuit-level 2-port configurations already provided by the memory compilers of chip-manufacturing vendors. 

Further, it has been shown that Algorithmic multi-ported memories are more efficient, in terms of resource usage and performance, when compared with synthesizing the same multiport memory with registers and has similar performance compared with a Circuit-level implementation \cite{msthesis}.

While, conventional approaches such as banking (which provides memory ports with conflicts), and mutli-pumping (which degrades the maximum external operating frequency), AMMs provide no-conflict based design that can operate at the maximum frequency. 

These properties of AMM can be leveraged in rapidly understanding the design space available in accelerator with several optimizations possible for high performance or EDP maximization objectives.

Most of the previous works of AMMs have explored its use cases on FPGAs \cite{malazgirt2014mipt,malazgirt2014application}, but the limited resource on FPGA constraints the full potential of their design space exploration. The limited number of BRAMs, F/F, slice LUTs are unable to explore important design factors of AMM such as the number of ports, memory depth and banking structures important in high-performance accelerators.

The contributions of our work are :
\begin{itemize}
    \item We extend the design space exploration for memories in Application-specific accelerators for high performance (in terms of execution time) by utilizing Algorithmic Multi-port Memories.
    \item We synthesize scratchpad and cache-memory AMM designs in different memory cells, port configurations and memory depth to improve design space exploration in Pre-RTL Accelerators. Then we schedule the multi-port memory instructions in the Aladdin framework by extracting parallelism from the algorithm trace.
    \item The efficient use of AMMs (as true multiport memories) compared to banking structures for high-performance design is dependent upon the spatial memory access-patterns of the application. We empirically depict this on the MachSuite benchmark and show that AMMs work well for high-performance design requirements, for applications with low spatial locality ($<$ 0.3) in memory access patterns.
\end{itemize}

\begin{figure}[!thpb]
    \centering
    \includegraphics[scale=0.5]{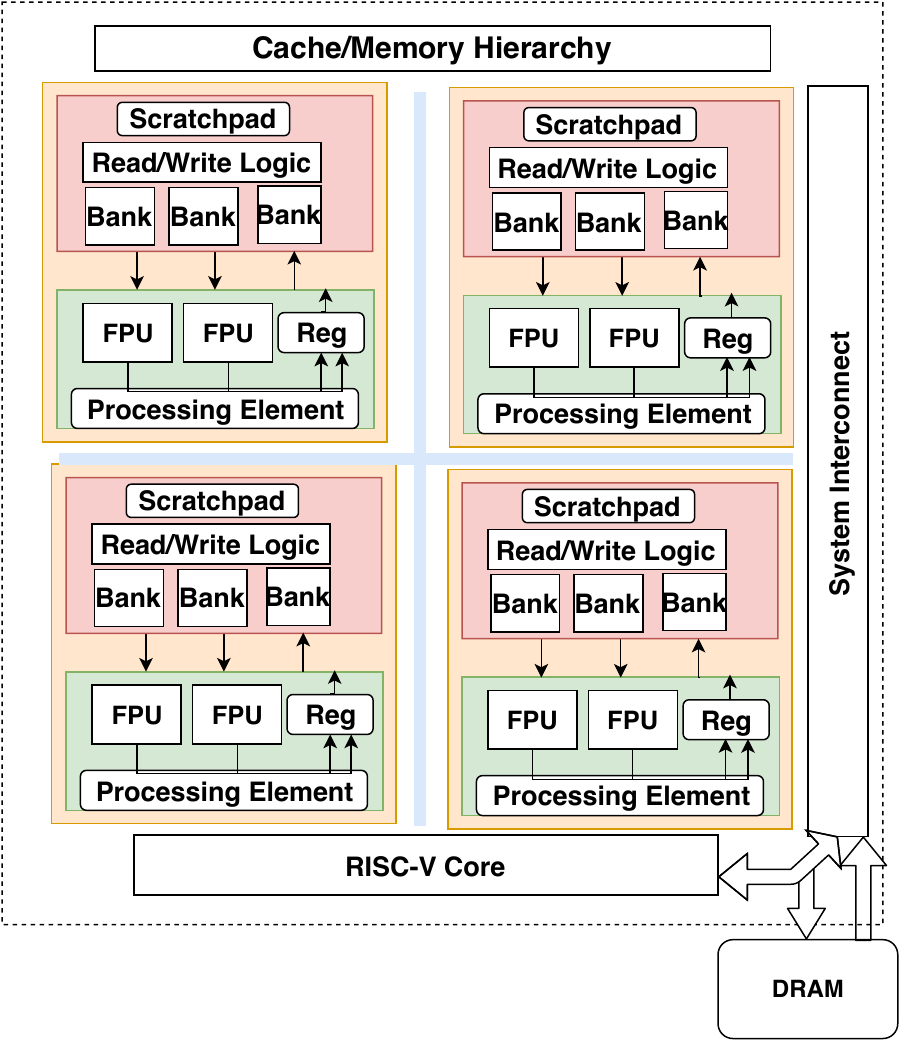}
    \caption{Accelerator-Based SoC with Multiport Banking-based Scratchpad Memory.}
    \label{fig:my_label}
\end{figure}
\section{Algorithmic Multi-port Memories}
The two-prominent approaches in AMMs include Non-table-based and Table-based.

\subsection{Non-Table-Based Approaches}
\begin{figure}[!thpb]
    \centering
    \includegraphics[scale=0.35]{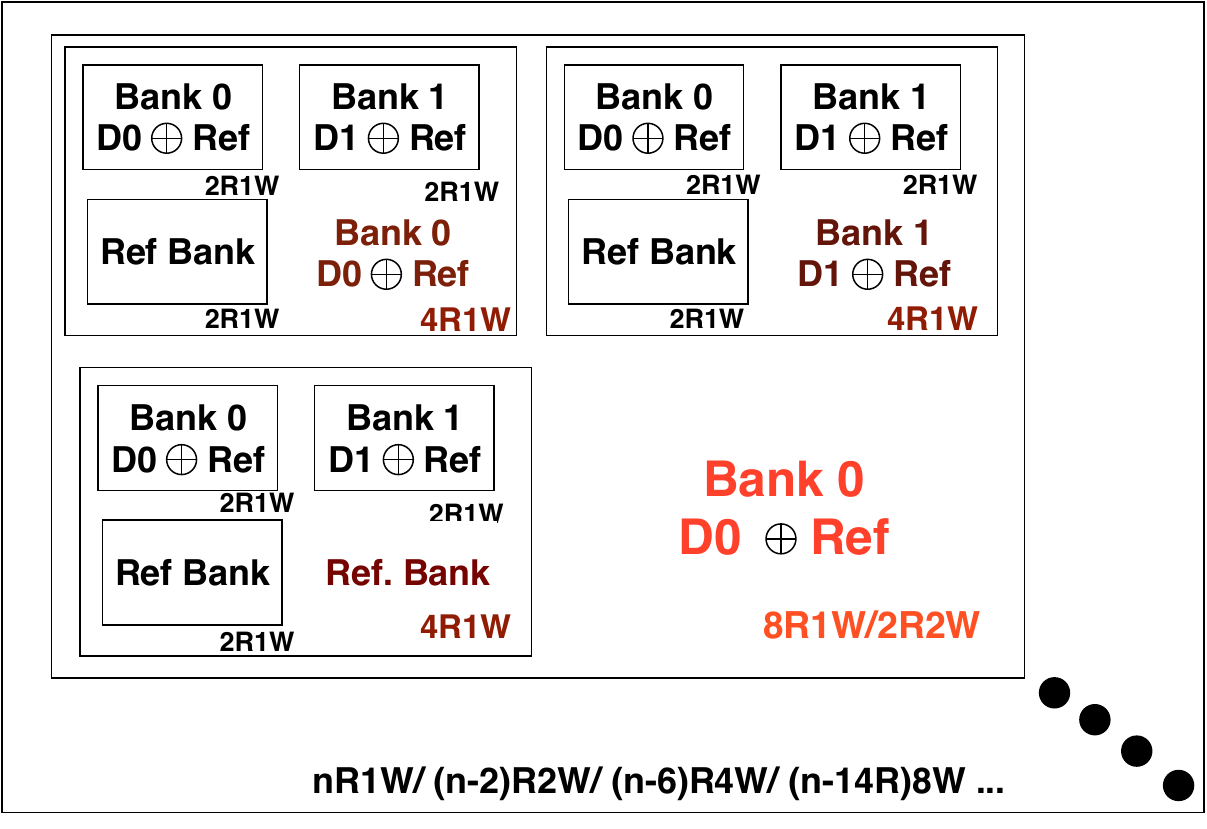}
    \caption{Flow of Increase of Read/Write Ports in HB-NTX-RdWr}
    \label{fig:my_label}
\end{figure}
The Non-table XOR-based approach (NTX-Rd/ NTX-Wr) proposed in , store xor-ed parity data in multiple banks to support multiple reads and enable multiple writes. It was developed further in HB-NTX-RdWr\cite{shahrouzi2017efficient} which can scale the number of ports with a systematic flow.\\
\textbf{H-NTX-Rd :} H-NTX-Rd follows the encoding scheme as, Bank0 stores Data0 directly, Bank 1 stores Data1 and Reference Bank stores D0 $\oplus$ D1. In case 2 Reads are directed to the same Bank, say Bank0, then the second read at offset i, can be retrieved as, Bank1[i] $\oplus$ Ref[i].\\
\textbf{B-NTX-Wr :} B-NTX-Wr follows the encoding scheme as, Bank0 stores Data0 $\oplus$ Ref, and Bank1 stores Data1 $\oplus$ Ref. In the Non-Conflict case, say two writes W0[i] and W1[j], are in different Banks, then data is stored as D0[i] = W0[i] $\oplus$ Ref[i] and D1[j] = W1[j] $\oplus$ Ref[j]. In the Conflict-based requests, say both W0 and W1 at Bank 0, then data is stored as, D0[i] = W0[i] $\oplus$ Ref[i], T = D1[j] $\oplus$ Ref[j], Ref[j] = W1[j] $\oplus$ D0[j] and D1[j] = Ref[j] $\oplus$ T. \\
\textbf{HB-NTX-RdWr :} HB-NTX-RdWr (shown in Fig. 2) adds reads ports following the scheme of H-NTX-Rd and write ports following the B-NTX-Wr write approach. Say, for building a 2R2W memory, all the banks should be made 4R1W following the H-NTX-Rd and the converted to 2R2W following the B-NTX-Wr method (Total read ports reduce because each read will access all the Banks and Each write will access it's own bank and the reference bank).\\
\subsection{Table-Based Approaches}
Table-based approaches use multiple memory modules to support multiple accesses and use lookup tables to avoid module conflict and track the most up-to-date values. 

The live value table (LVT) approach utilizes LUT to track the most updated location of the stored data. Read requests query the table to access data at the correct memory location. The multiple read requests are handled by replicating memory banks, and multiple write requests are supported by the LVT.

Usually, Non-table-based AMMs have shorter latencies when compared with table-based designs. Table-based AMMs pose smaller area and lower power consumption than non-table-based AMMs. 
Other approaches include Table-Based Remap Table(TBRemap), (TBLVT-HB-NTX-RdWr) and enhancing Table-Based techniques with Reduce Lookup Tables. \cite{laforest2012multi,laforest2010efficient,lai2016efficient,lai2017efficient,Abdelhadi:2014:MMS:2554688.2554773,shahrouzi2017efficient,lin2015bram}



\section{Experimental Setup}
\subsection{Synthesis} 
All the AMM designs discussed above (both Table and Non-table Based), were synthesized and the best performing configurations (of a given read-write port and port depth) were chosen to be incorporated in the power, area and latency models in the Aladdin Framework. 
The write-path and read-path logic (in Verilog HDL) was synthesized in Synopsys Design Compiler at UMC 45nm. CACTI is used to estimate the performance of different SRAM modules. By combining the synthesis results of read-path and write-path logic, and estimation from CACTI (SRAM) we can evaluate the overall performance and cost of an AMM design.

\subsection{Flow of Analyzing}
In this section, we describe the flow of analysis in the Aladdin Framework \cite{shao2014aladdin} and the methodology used for incorporating multi-port memory access. 
The Aladdin Framework transforms the source code for the accelerators to output configuration results including clock speed, the number of cycles, components in the synthesized design, power and area breakdown of each component and its total utilization. The code is compiled to the LLMV IR and a Data Dependency Graph (DDG) is extracted from it. Graph Transformations are applied according to the user-defined configurations, where multiple memory banks can be specified by array partitioning, multi-issue ALUs may be constructed by loop unrolling and flattening etc. Further, caches and their structure can be configured and registers and scratchpad memory can be utilized to club or unroll memory elements.
From these specifications, a cycle-accurate execution takes place, from the resources calculated earlier and power, area and execution time are determined. \\
For utilizing the multiport memories, we add our own power, area and latency results from the synthesis as described above. 

\subsection{Scheduling for Multiport Memories}
The cycle-accurate simulator schedules the data flow graph, which is represented as a DAG that extracts memory level parallelism of memory instructions and instruction-level parallelism of ALU operations. 

The DAG allows multiple accesses and the scheduler then issues the number of accesses requested, accordingly from the read-write port configurations and port width defined by the user. 

\begin{figure}[!thpb]
    \centering
    \includegraphics[scale=0.50]{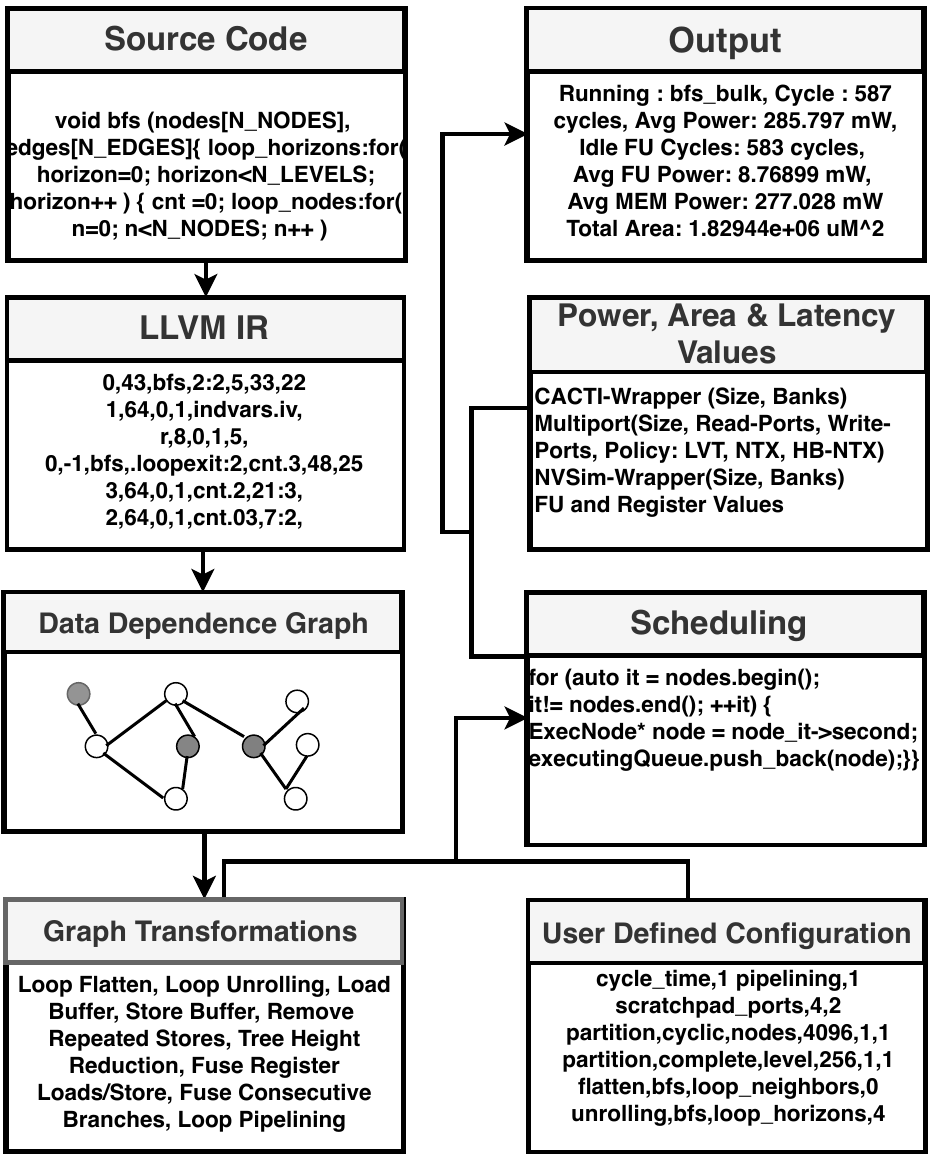}
    \caption{Flow of Methodology Used from Source Code to Output used in Aladdin Framework.}
    \label{fig:my_label}
\end{figure}

\begin{figure*}[!thpb]
    \centering
    \includegraphics[scale=0.45]{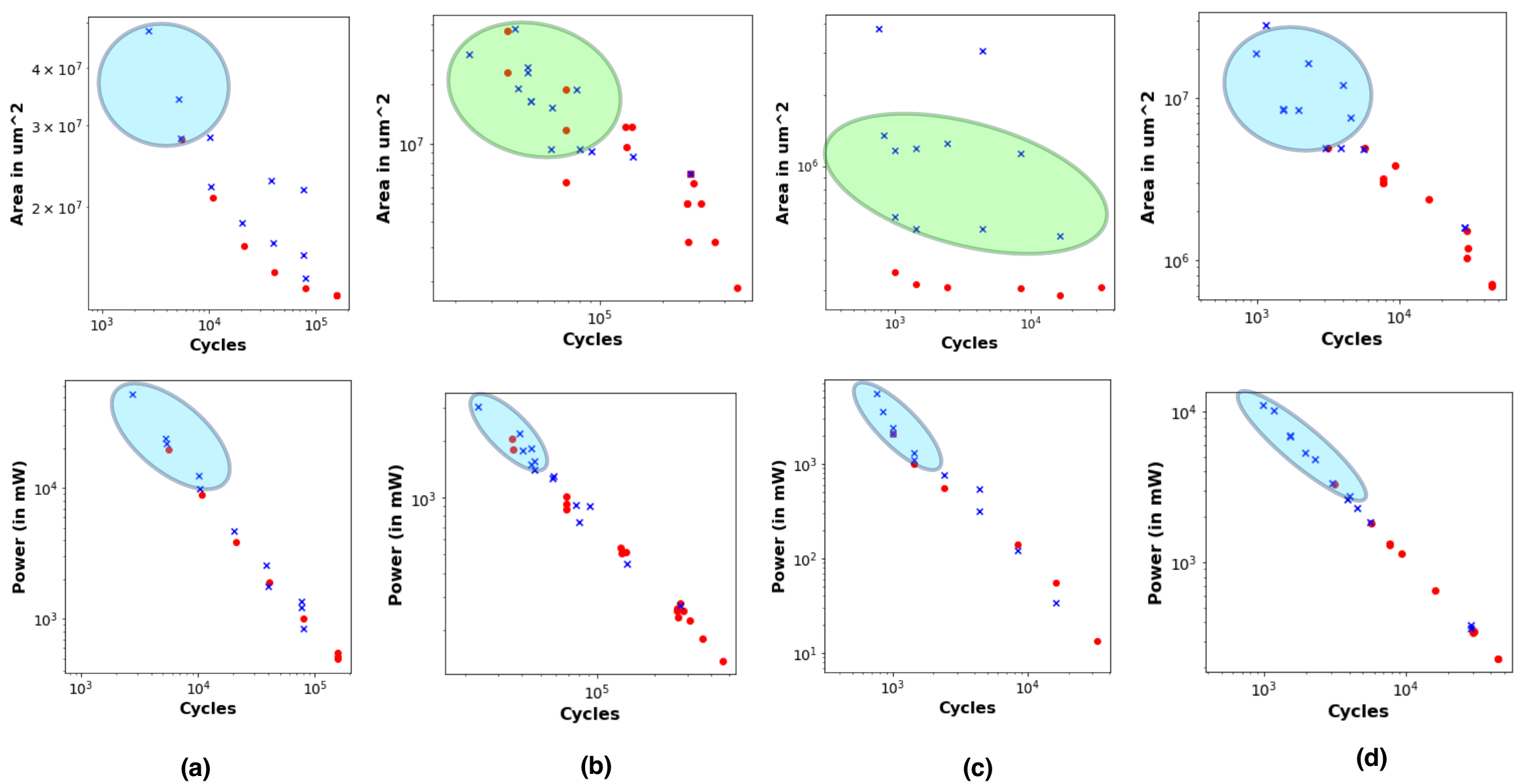}
        \caption{Area-Cycles and Power-Cycles Design Space Exploration on Benchmarks (a)FFT-Strided (b)GEMM-NCUBED (c)KMP (d)MD-KNN. Increase in design space is shaded blue.}
    \label{fig:my_label}
\end{figure*}
\section{Discussion}
\subsection{Design Space Exploration of AMM}
In the design space exploration of any accelerator design, different compositions are possible by loop-unrolling, array-partitioning, changing word-size and number of read and write ports. We use a sweep of such compositions, in the implemented Mem-Aladdin Framework.

We analyze using the MachSuite benchmark which includes both compute-intensive benchmarks such as, Stencil, where arithmetic and functional units often dominate execution time and power, and memory-bound workloads, e.g., Sort, to give modelling capabilities across multiple dimensions. 

From the benchmarks suites, the four benchmarks for discussion (FFT-Strided, GEMM-NCUBED, KMP, MD-KNN) are chosen accordingly with varying spatial locality (discussed ahead) and the memory size requirements in execution. 

In terms of the memory configurations, a wide range of designs are possible by partitioning large arrays into scratchpad memories allowing parallel access. 

The arrays which have single-stride access can be partitioned cyclically. The parallel access with a number of ALU/FPU can allow faster execution but increase the power consumption and area of the design. However, for different strides, the accesses will be different. The array-partitioning based design based exploration is dependent on the spatial locality of the array in the algorithm. Multiport Memory designs provide non-conflict based access and independent of the stride access. 

Figure 4 shows the design space exploration by varying these parameters for four different benchmarks, chosen to demonstrate and understand the benefits of algorithmic multiport memory. The designs utilizing multiport memory are marked in blue and with single port-memories are shown in red. The design space can be increased (shaded in blue) by using multiport memory creating high-performance accelerators that were not possible earlier.
\begin{figure}[!thpb]
    \centering
    \includegraphics[scale=0.4]{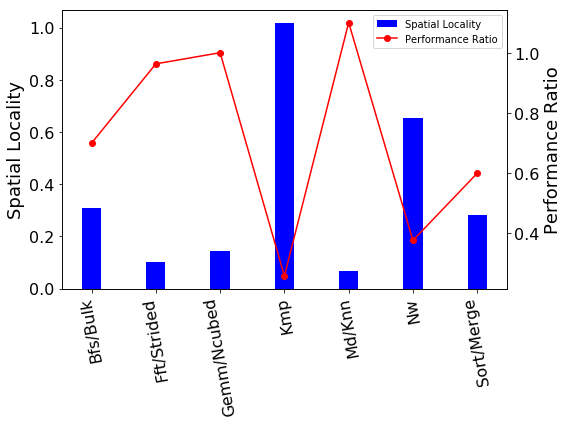}
    \caption{Spatial Locality and Performance Ratio (in terms of Area Requirements) (higher is better) comparison for various benchmarks.}
    \label{fig:my_label}
\end{figure}
\subsection{Spatial Locality}
Both stride and length of the word are directly correlated with performance in array-partitioned based designs. Algorithms with high-stride will perform poorly in speed due to high bank-conflicts and having larger word-size of each element will increase the area and power consumption of the design.
Comparing Array-partitioning based banking and algorithmic multiport memory is important to understand the need of using the later in chip designs.

Spatial locality measures where a program’s memory accesses occur in relation to each other. To summarize the spatial locality across MachSuite, we use a simple metric defined by Weinberg et al. \cite{weinberg2005quantifying}  :
\begin{equation}
L_{spatial} =
\sum_{stride=1}^{stride<\infty} \frac{P(stride)}{stride}
\end{equation}
where stride is the difference between consecutive address elements referenced and length is the number of of elements with same stride in a load/store instruction. 

The spatial locality on the MachSuite benchmark is shown in Fig 5. From equation 1, programs with stride-one code have high spatial locality. 
Stride-one code is available in byte-oriented programs like KMP and AES, where double-precision programs (e.g. FFT-Strided) have a minimum stride distance of 8 bytes. 

From Fig 4, it is seen that the performance of algorithmic multi-port memory is similar to having a large number of small partitions of the array.

For most of the benchmarks, the Power Consumption vs Execution time trade-off for both AMM and banking-based designs is similar. It can be due to the fact banking-based design use multiple smaller memory sizes and are slower due to conflicts which can easily be offset by the using larger memory AMM design. For similar speed both the approaches take usually similar number of cycles.  

\subsection{Qunatifying Correlation between AMM performance and Spatial Locality}
The performance in AMM is correlated here to spatial locality of the benchmarks.  Since KMP uses 1-bit word access with stride 1, it has very less number of bank conflicts in array-partitioning so a true-multiport design isn't necessary. Hence, in the area required vs the number of cycles, the area required by AMM is higher in KMP (shaded green) whereas it is nearly equal in other cases. The design space using AMM uses more area, hence performs worse than array-partitioning design. 

In Gemm-Ncubed design in Fig 4 (c), the AMM performs better with correct word size and read/write port configuration. The spatial locality of Gemm is low because of higher word-size since computation is done in floating-point numbers.

Since only the area required for particular execution time varies significantly in AMM based designs vs banking-based structures, we compare for area-requirements only. This comparison can be represented as a single metric over the entire design space, by taking the geometric mean over the observed points. More Specifically, 

Performance Ratio = $({(a_1*a_2....*a_n)}/{(b_1*b_2....*b_n)})^{1/n}$,
where ${a_i}$ is the area-point of banking-structure, ${b_i}$ is the area-point of AMM at similar execution times.

Further, this analysis can only be done for benchmarks with high number of memory accesses compared to computation operations. Hence, Pre-RTL design of only a few benchmarks can be chosen to demonstrate this behaviour. 

The correlation of performance of AMM-based design (taking banking-structures as baseline) with Spatial Locality is also shown in Fig 5. 

Hence, AMMs can be effectively utilized to provide non-conflict multi-port memory access for designing high-performance accelerators of applications with low spatial locality ($<$ 0.3) in memory access patterns.

\section{Conclusions}
In this work, we present a method to incorporate AMM based scratchpad and cache-memory multi-port configuration to improve design space exploration in Pre-RTL Accelerators. We describe a method which schedules multi-port memory instructions in the Aladdin framework by extracting parallelism from the algorithm trace. We synthesize different AMM designs, memory depth, port configurations, banking structures, building memory cells.  From our analysis on applications from the MachSuite (accelerator benchmark suite), we gain insight into the use of AMMs (as true multiport memories) for high performance in applications with low spatial locality in memory access patterns which is quantified by empirical observations. 


\bibliographystyle{IEEEtran}
\bibliography{ref}
\end{document}